\documentclass[prb, twocolumn, superscriptaddress, amsmath, amssymb, 10pt]{revtex4-2}

\usepackage{float, graphicx}
\usepackage[dvipsnames]{xcolor}
\usepackage[caption=false]{subfig}
\usepackage{physics}
\usepackage{epstopdf}
\usepackage{hyperref}
\usepackage{lipsum}
\hypersetup{linkcolor=blue, citecolor=blue, filecolor=blue,urlcolor=blue, colorlinks=true}
\usepackage{soul} 

\renewcommand{\v}[1]{\ensuremath{\boldsymbol{\mathbf{#1}}}}

\newcommand{\UIUCPHYS}[0]{Department of Physics, University of Illinois at Urbana-Champaign, Urbana, IL 61801, USA}
\newcommand{\UICPHYS}[0]{Department of Physics, University of Illinois at Chicago, Chicago, IL 60607, USA}

\begin{document}

\title{Competing Higher Order Topological Superconducting Phases in Triangular Lattice Magnet-Superconductor Hybrid Systems}

\author{Ka Ho Wong}\affiliation{\UICPHYS}
\author{Jacopo Gliozzi}\affiliation{\UIUCPHYS}
\author{Mark R. Hirsbrunner}\affiliation{\UIUCPHYS}
\author{Arbaz Malik}\affiliation{\UICPHYS}
\author{Barry Bradlyn}\affiliation{\UIUCPHYS}
\author{Taylor L. Hughes}\affiliation{\UIUCPHYS}
\author{Dirk K. Morr}\affiliation{\UICPHYS}

\begin{abstract}
We demonstrate that a plethora of higher order topological phases emerge in magnet-superconductor hybrid (MSH) systems through the interplay of a stacked magnetic structure and an underlying triangular surface lattice; the latter being of great current experimental interest. Such lattices offer the ability to create three main types of edge terminations -- called $x$-, $y$- and $y^\prime$-edges -- of MSH islands that, in turn, give rise to a complex phase diagrams exhibiting various regions of HOTSC phases. We identify the single adatom chain, as well as a pair of adjacent adatom chains (called a double-chain), as the basic building blocks for the emergence of HOTSC phases. Of particular interest are those HOTSC phase which arise from a competition between the topology of single and double-chain blocks, which are absent for square latices.

\end{abstract}

\maketitle
\nopagebreak
\section{Introduction}
Majorana zero modes (MZMs) that emerge in topological superconductors have non-Abelian statistics and topological protection against disorder and decoherence, which can be utilized for fault-tolerant quantum computing~\cite{Nayak2008}. Among the most promising material platforms for the creation of topological superconducting phases are semiconductor-superconductor hybrid systems \cite{Lutchyn2010, Mourik2012,Aasen2016, Lutchyn2018}, as well as magnet-superconductor hybrid (MSH) systems \cite{Nadj-Perge2014,Ruby2015,Pawlak2016,Kim2018,Menard2017,Palacio-Morales2019,Kezilebieke2020,Bazarnik2023}. In the latter, magnetic adatoms are deposited on the surface of $s$-wave superconductors to form one-dimensional (1D) chains \cite{Nadj-Perge2014,Ruby2015,Pawlak2016,Kim2018} or two-dimensional (2D) islands \cite{Menard2017,Palacio-Morales2019,Kezilebieke2020,Bazarnik2023}. Experiments have shown evidence for strong topological superconducting phases in 2D ferromagnetic MSH systems \cite{Menard2017,Palacio-Morales2019,Kezilebieke2020,Bazarnik2023}, and recent scanning tunneling spectroscopy experiments have also revealed evidence for topological nodal superconductivity in antiferromagnetic (AFM) MSH systems \cite{Bazarnik2023}. A series of theoretical studies have also predicted the existence of strong topological phases in MSH systems having skyrmionic~\cite{Mascot2021}, checkerboard \cite{Kieu2023}, or 3{\bf Q}-magnetic structures~\cite{Bedow2020}.

Recently, higher order topological superconducting (HOTSC) phases have been predicted in more complex AFM \cite{Zhang2019} or stacked \cite{Wong2023} magnetic structures. The predicted HOTSC phases realize extrinsic, boundary-obstructed higher order topology~\cite{benalcazar2017prb,benalcazar2017science,Geier2018,Khalaf2021}, which yields phenomenology that depends on the details of the boundary geometry and terminations of finite sized MSH systems. Since the types of possible edge terminations in MSH systems strongly depend on the underlying lattice structure of the (superconducting) surface on which the magnetic adatoms are patterned, the question immediately arises of how the square-lattice results of Ref. \onlinecite{Wong2023} can be generalized to other experimentally relevant lattices such as triangular lattice Re \cite{Palacio-Morales2019} surfaces or two-dimensional NiSe$_2$/CrBr$_3$ heterostructures~\cite{Kezilebieke2020}, or the projected Nb bcc-lattice \cite{Bazarnik2023}, can give rise to novel unique new mechanisms of HOTSC phases.

In this article, we address this question by investigating the emergence of HOTSC phases in MSH systems that possess a triangular (superconducting) surface lattice. Such a lattice allows for three primary types of edge terminations --which we call $x$-, $y$- and $y^\prime$-edges -- of finite MSH island samples. The interplay between the extrinsic higher order topology and the boundary geometry gives rise to a complex phase diagram exhibiting distinct regions of HOTSC phases. In particular, we show that, in contrast to MSH islands on a square lattice \cite{Wong2023}, for a triangular lattice, HOTSCs can arise from coupling arrays of single adatom chains, as well as adjacent pairs of adatom chains (called double-chains). This, in turn, is the source of the rich phenomenology of HOTSC phases, all of which can be determined from the patterns of dimerized couplings between the single or double-chain building blocks. Of particular interest are the HOTSC phases that arise from a competition between the topology of single and double-chain blocks, which are absent for square latices. Finally, we show that the ability to create MSH islands of triangular, trapezoidal, or hexagonal shape provides an unprecedented tool to manipulate the existence and location of Majorana zero modes. Importantly, advances in atomic manipulation techniques \cite{Kim2018} have made the quantum engineering of the MSH systems we describe, and hence the resulting HOTSC phases, a real possibility for the near future.

\section{Magnetic-Superconductor Hybrid Model}
\begin{figure}[ht!]
\centering
\includegraphics[width=0.9\linewidth]{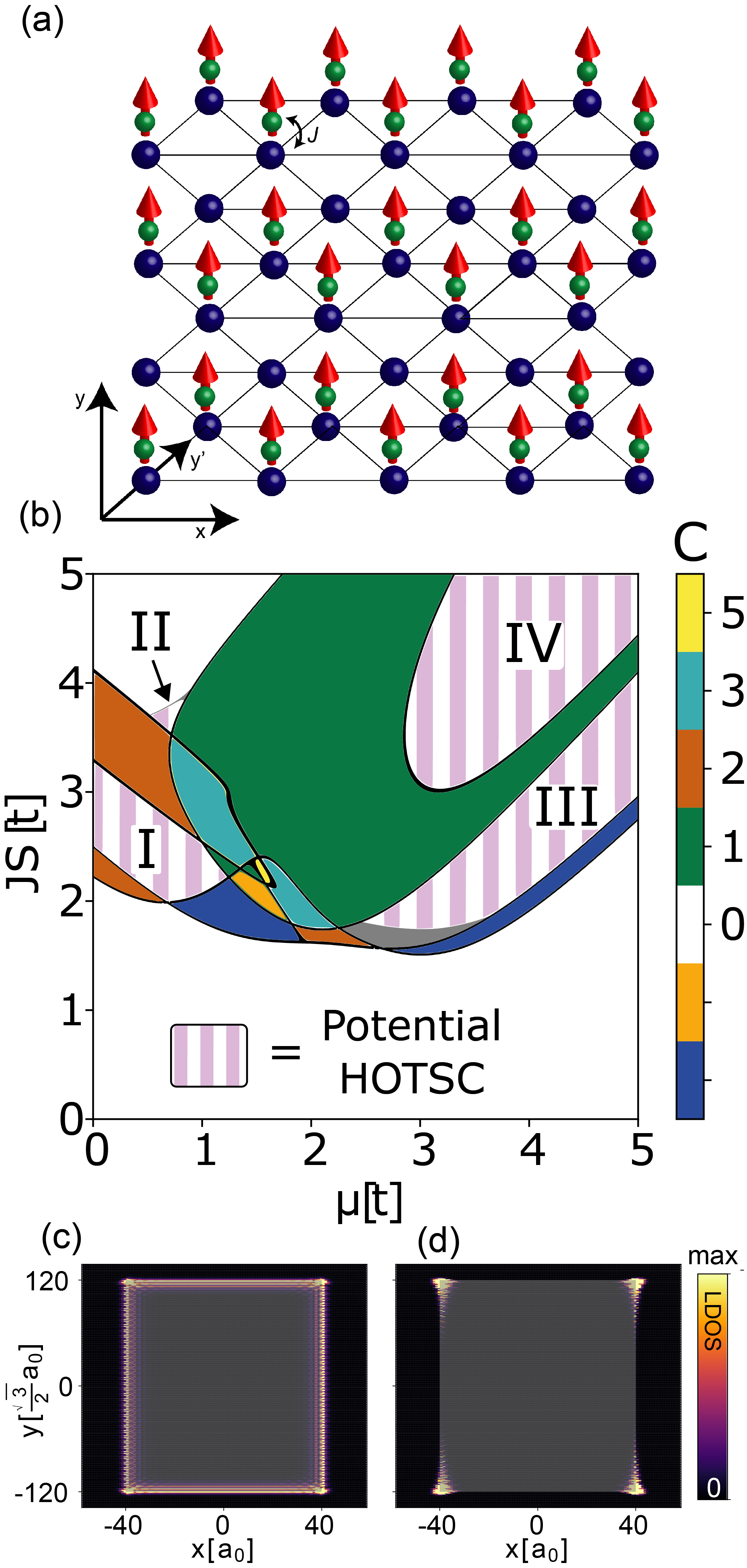}
  \caption{(a) Schematic view of the MSH system with double-chains of magnetic adatoms: dark blue circles represent sites of the superconducting substrate lattice and green circles with red arrows denote the magnetic adatoms. (b) Topological phase diagram of the MSH system, with strong topological phases colored according to their Chern number. Regions with potential HOTSC phases are shown in a striped pattern. Edge gap closings are shown in grey. Zero-energy LDOS for a finite-sized island in the (c) strong topological $C=1$ phase reflecting the existence of a chiral edge modes, and (d) in a HOTSC phase, showing the presence of characteristic Majorana corner modes.}\label{fig:1}
\end{figure}

Topological superconductivity in two-dimensional (2D) magnet-superconductor hybrid systems arises from the interplay of a broken time-reversal symmetry, a Rashba spin-orbit coupling, and a hard ($s$-wave) superconducting gap \cite{Ron2015,Li2016,Rachel2017}. A Rashba spin-orbit coupling is naturally induced by the broken inversion symmetry on the surface of the superconducting substrate, and placing magnetic adatoms on the surface breaks time-reversal symmetry. We therefore consider the following Hamiltonian to describe the MSH system~\cite{Ron2015,Li2016,Rachel2017}:
\begin{equation}
    \begin{aligned}
        \mathcal{H} =& -t \sum_{\v{r}, \v{\delta},  \alpha} c^\dagger_{\v{r}, \alpha} c_{\v{r}+\v{\delta}, \alpha} - \mu \sum_{\v{r}, \alpha} c^\dagger_{\v{r}, \alpha} c_{\v{r}, \alpha} \\
        &+ i \lambda \sum_{\substack{ \v{r}, \v{\delta} \\ \alpha, \beta}} c^\dagger_{\v{r}, \alpha} \left(  \left[\v{\delta} \times \v{\sigma} \right] \cdot \hat{z} \right)_{\alpha, \beta}  c_{\v{r} + \v{\delta}, \beta} \\
        &+ \Delta \sum_{\v{r}} \left( c^\dagger_{\v{r}, \uparrow} c^\dagger_{\v{r}, \downarrow} + c_{\v{r}, \downarrow} c_{\v{r}, \uparrow} \right) \\
        &+ J {\sum^\prime_{\substack{\v{R} \\ \alpha, \beta}}} c^\dagger_{\v{R}, \alpha} \left[ \v{S}_{\bf R} \cdot \v{\sigma} \right]_{\alpha,\beta} c_{\v{R}, \beta} \; .
    \label{eq:H}
    \end{aligned}
\end{equation}
Here $c^\dag_{\v{r}, \alpha}$ is the creation operator for an electron of spin-$\alpha$ at site $\v{r}$ on the two-dimensional triangular lattice, $-t$ is the hopping strength between nearest-neighbor sites (which are separated by a lattice vector $\v{\delta}$), $\mu$ is the chemical potential, $\lambda$ is the strength of the Rashba spin-orbit coupling, $\v{\sigma}$ is a vector of spin-1/2 Pauli matrices, $\Delta$ is the $s$-wave superconducting order parameter, and $J$ is the magnetic exchange interaction between the adatom spin $\v{S}_{\v{R}}$ and the conduction electrons at site $\v{R}$. As such, the primed sum runs over only the sites of the triangular lattice that are decorated with an adatom, as shown in Fig.~\ref{fig:1}{a}. Furthermore, we treat the adatoms as classical spins since Kondo screening is suppressed by the hard superconducting gap~\cite{Balatsky2006,Heinrich2018}. For our calculations we consider adatoms that are ferromagnetically aligned out of the plane of the superconducting substrate, as shown in Fig.~\ref{fig:1}(a), and take the superconducting pairing and Rashba spin-orbit coupling to be of strength $\Delta = 1.2t$ and $\lambda= 0.8t$, respectively. While this value of $\Delta$ appears relatively large, we note that due the presence of the magnetic adatoms, the actual gap in the electronic spectrum is significantly smaller, and varies across the phase diagram. The specific value of $\Delta$ was chosen for illustrative convenience in our discussion of the topological phase diagram and to minimize finite size effects in the LDOS calculations for the spatially localized MZMs. However, we find that the  qualitative nature of our results, and specifically the existence of multiple HOTSC phases, is independent of the specific value of $\Delta$ (see Supplementary Material Sec.I).

As a Bogoliubov-de-Gennes (BdG) superconductor Hamiltonian without time-reversal symmetry, the model in Eq.~\eqref{eq:H} belongs to class D of the Altland-Zirnbauer classification of free fermion band topology~\cite{altlandzirnbauer,ryu2010topological,kitaev2009periodic}. Two-dimensional strong topological phases in class D are classified by the Chern number, an integer invariant defined as
\begin{equation}\label{Chern}
\begin{aligned}
C &= \frac{1}{2\pi i} \int_{\text{B.Z.}} d^2k \, \text{Tr} \, (P_{\v{k}} [\partial_{k_x} P_{\v{k}}, \partial_{k_y} P_{\v{k}}]), \\
P_{\v{k}} &= \sum_{E_n(\v{k})<0} \ket{\psi_n(\v{k})} \bra{\psi_n(\v{k})},
\end{aligned}
\end{equation}
where $E_n(\v{k})$ and $\ket{\psi_n(\v{k})}$ are, respectively, the single-particle energies and states of the BdG Hamiltonian, $\v{k}$ is the crystal momentum, and both the trace and the index $n$ run over spin, magnetic-sublattice, and Nambu degrees of freedom.

 While the presence of a non-zero spin-orbit coupling and a uniform ferromagnetic structure of the adatoms is sufficient to induce strong topological superconductivity in MSH systems, the emergence of a HOTSC phase requires additional ingredients. In particular, it was previously shown~\cite{Wong2023} that the dimerized couplings between magnetic adatoms resulting from a spatial modulation in their arrangement on the surface can give rise to extrinsic, boundary-obstructed HOTSC phases. Such phases are separated from trivial phases by transitions at which gap closings occur on some, but not all, edges of the MSH system~\cite{Geier2018, Khalaf2021}. Such phases are less stable than intrinsic HOTSCs, as the MSH system can be trivialized by purely edge perturbations. However, this boundary sensitivity can also be a feature since it enables heightened tunability of the MZMs that we discuss below.

 Following Ref.~\onlinecite{Wong2023}, we consider MSH structures in which adatom chains oriented along the $x$-direction are arrayed and patterned in the $y$-direction, and restrict our attention to repeating patterns of two adjacent adatom chains followed by an empty row of superconducting substrate [see Fig.~\ref{fig:1}(a)].
 As was previously shown\cite{Wong2023}, the individual 1D chains of adatoms can enter a 1D strong topological superconductor phase having MZM end states and a non-trivial $\mathbb{Z}_2$ bulk invariant (Altland-Zirnbauer class D)\cite{Kitaev2001}. Uniform arrays of such chains in 2D can produce weak TSC phases that possess gapless edge states on certain edges that are protected by translation symmetry in the stacking direction~\cite{seroussi2014topological}. We show below that breaking this translation symmetry via dimerization gaps out the weak TSC boundary modes, leaving unpaired MZM corner modes. Hence, dimerized couplings of a stack of these chains can produce the HOTSC phases~\cite{Wong2023}.

\begin{figure}[ht!]
\centering
\includegraphics[width=0.9\linewidth]{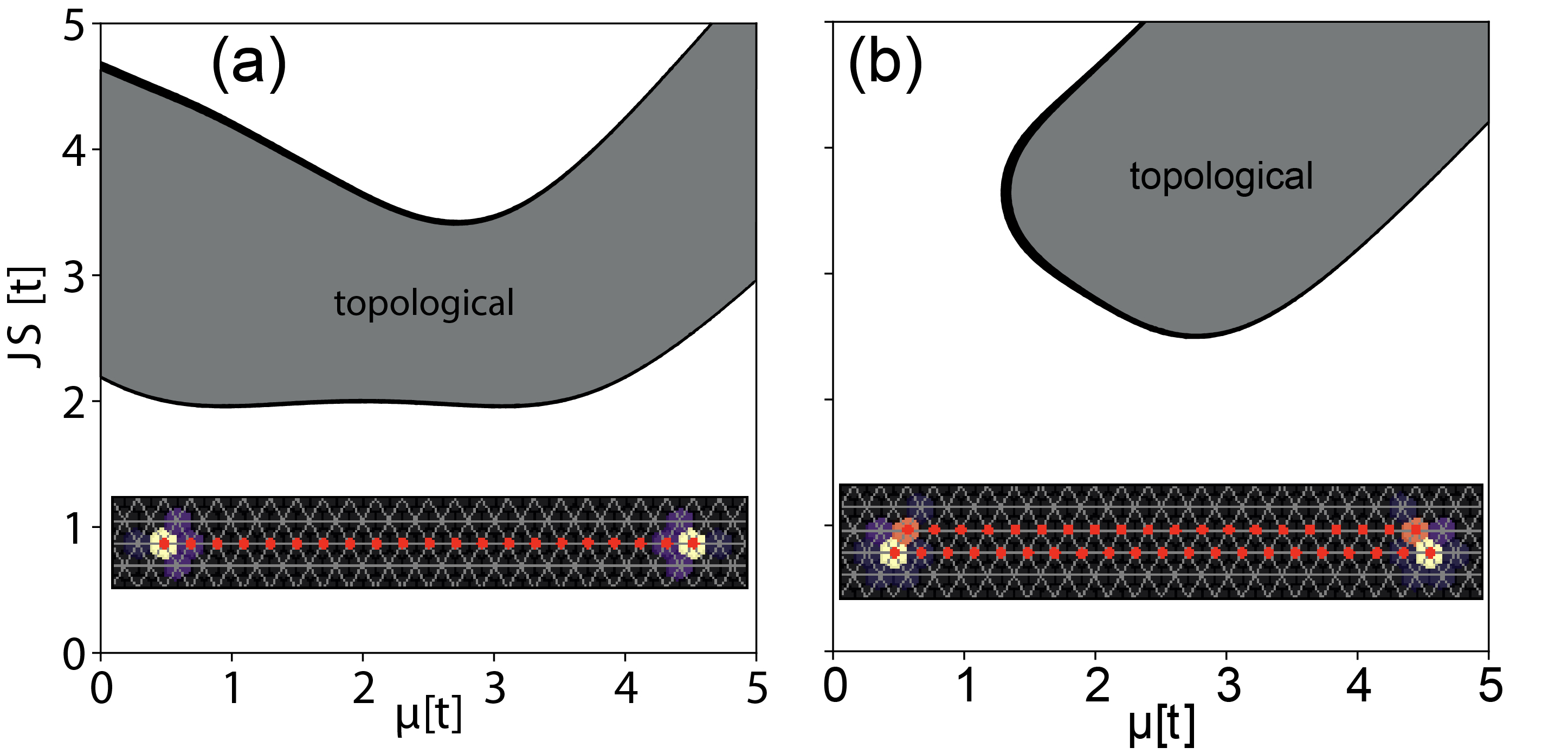}
  \caption{Phase diagrams of (a) an isolated single adatom chain and (b) an isolated double adatom chain. Dark gray and white regions indicate topological and trivial phases, respectively. Insets show the corresponding zero-energy LDOS of single and double chains in the topological regime.
  }\label{fig:single}
\end{figure}

\section{Topological Phases of the MSH Model}
In Fig.~\ref{fig:1}(b) we present the topological phase diagram of the MSH model. We will now describe the variety of topological phases that appear. Before discussing the HOTSC phases of the MSH model described above, we briefly review the salient features of the strong topological phases. The strong phases are characterized by a non-vanishing Chern number $C$, and phase transitions where the Chern number changes are accompanied by bulk gap closings. In the phase diagram in Fig.~\ref{fig:1}(b) we observe strong TSC phases having Chern numbers ranging from $C=-2$ to $C=5$. The solid black lines bounding these phases correspond to bulk gap closings at high symmetry points in the Brillouin zone. All finite size MSH islands in any of these strong TSC phases possesses chiral edge modes along all edges, irrespective of edge termination or the presence of disorder that preserves the bulk gap. The edge states generate a non-vanishing zero-energy local density of states along the edges, as shown  Fig.~\ref{fig:1}(c).

In contrast, a HOTSC phase is identified by the emergence of MZMs at the corners of finite-size islands -- so-called Majorana corer modes -- as shown in Fig.~\ref{fig:1}(d). Contrary to strong TSC phases, the existence of HOTSC phases, and thus the presence or absence of Majorana corner modes, sensitively depends on the details of the edge termination. As such, we must carefully consider the edges that can be realized in a finite-size island of magnetic adatoms due to the underlying triangular lattice of the superconducting surface.

To address this we note that MSH islands of magnetic adatoms placed on a triangular lattice can possess three different types of edges, one along each of the $x$-, $y$-, and $y^\prime$-directions, as shown in Fig.~\ref{fig:1}(a). We refer to these types of edges as $x$-, $y$-, and $y^\prime$-edges, respectively. While the $x$-, and $y^\prime$-edges are straight edges, the $y$-edge is formed by a zig-zag arrangement of magnetic adatoms, an arrangement that is instrumental in determining the mechanisms for the emergence of HOTSC phases, as we show below. Moreover, in contrast to strong TSC phases, HOTSC phases are bounded by gap closings on edges, rather than bulk gap closings. The regions in the phase diagram where such gap closings occur sensitively depend on the type of edge termination. Hence, to identify the edge transitions we must carry out a careful analysis of the phase diagram for each of the $x$-, $y$-, and $y^\prime$-edges. To this end, we consider the HOTSC phases of two different types of MSH islands: those with $x$- and $y$-edges  (Sec.~\ref{sec:xy-edges}),  and those with $x$- and $y^\prime$-edges (Sec.~\ref{sec:xyprime-edges}).

\begin{figure*}
  \includegraphics[width=\textwidth]{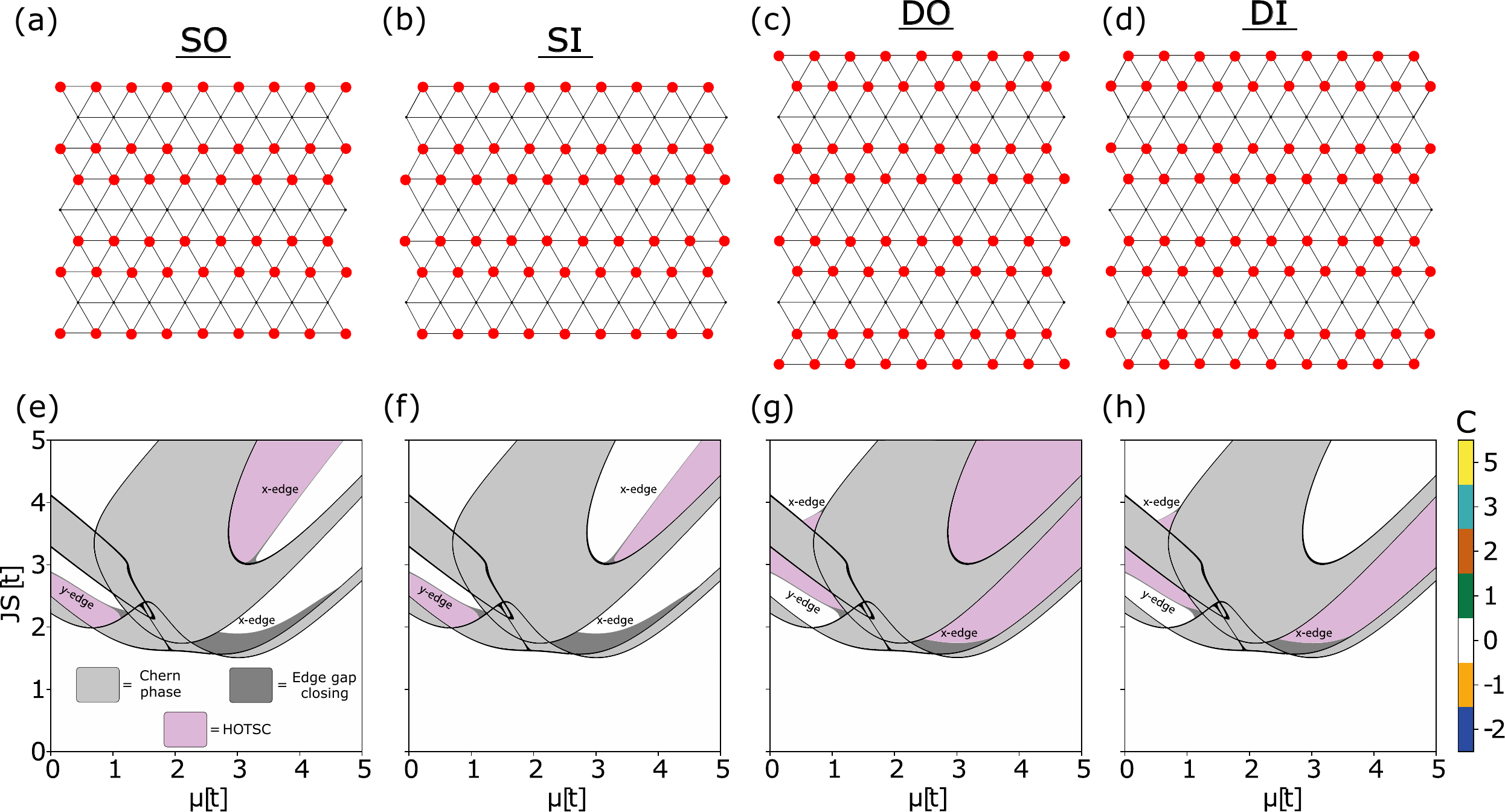}
  \caption{(a)-(d) Four different edge terminations of MSH islands with $x$- and $y$-edges, and (e) - (h) the corresponding phase diagrams: (a),(e) double chain termination, pointing out (DO), (b),(f) double chain termination, pointing in (DI), (c),(g) single chain termination, pointing out (SO), and (d),(h) single chain termination, pointing in (SI). HOTSC phases are denoted by pink regions, strong topological phases with non-zero Chern numbers by light grey regions, and edge gap closings by dark grey regions and lines. }\label{fig:2}
\end{figure*}

As we show below, the topology of the HOTSC phases for different MSH island geometries arises from either the topology of the single adatom chain, or that of a pair of coupled adatom chains located on adjacent rows that together form a non-trivial 1D TSC [called a double chain, see Fig.~\ref{fig:1}(a)]. The phase diagrams of (isolated) single and double chains possess large regions with 1D strong $\mathbb{Z}_2$ topological phases that are equivalent to the Kitaev chain~\cite{Kitaev2001}, as shown in Figs.~\ref{fig:single}(a) and (b), respectively. In these 1D strong TSC phases, the respective single or double chains host MZMs at their ends, as shown in the insets of Figs.~\ref{fig:single}(a) and (b) where we show the corresponding zero-energy LDOS. We hence expect that HOTSC phases of MSH islands can occur in all of regions of the phase diagram where either the single or double chain is topological, and where the MSH system does not exhibit a strong topological phase; such regions are marked as regions I-IV in Fig.~\ref{fig:1}(b). We will now show in detail how our expectation is borne out in the phase diagram of our model.

\subsection{HOTSC phases of MSH islands with $x$- and $y$-edges}
\label{sec:xy-edges}
While the phenomenological pattern of HOTSC phases emerging in MSH systems having different edge terminations is complex, we will now detail the general underlying mechanisms for these HOTSC phases and explicitly demonstrate how the HOTSC phases arise from the interplay between the stacking and coupling of the basic single or double-chain units described above.

We begin by considering MSH islands with $x$- and $y$-edges. For such an island geometry, several types of edge terminations are possible since the $x$-edge of the island can terminate either on a single adatom chain or on a double adatom chain. Furthermore, for each of these choices of $x$-edge terminations, the associated $y$-edges can point either \emph{in} or \emph{out} at the corners, as illustrated in Figs.~\ref{fig:2}(a)-(d). We refer to these four different terminations of rectangular islands as single out (SO) [Fig.~\ref{fig:2}(a)], single in (SI) [Fig.~\ref{fig:2}(b)], double out (DO) [Fig.~\ref{fig:2}(c)], or double in (DI) [Fig.~\ref{fig:2}(d)] terminations.
The corresponding phase diagrams in Figs.~\ref{fig:2}(e)-(h) reveal that HOTSC phases can occur in all regions I-IV [see Fig.~\ref{fig:1}(b)], depending on the edge termination. 

\begin{figure}[h!]
\centering
\includegraphics[width=\linewidth]{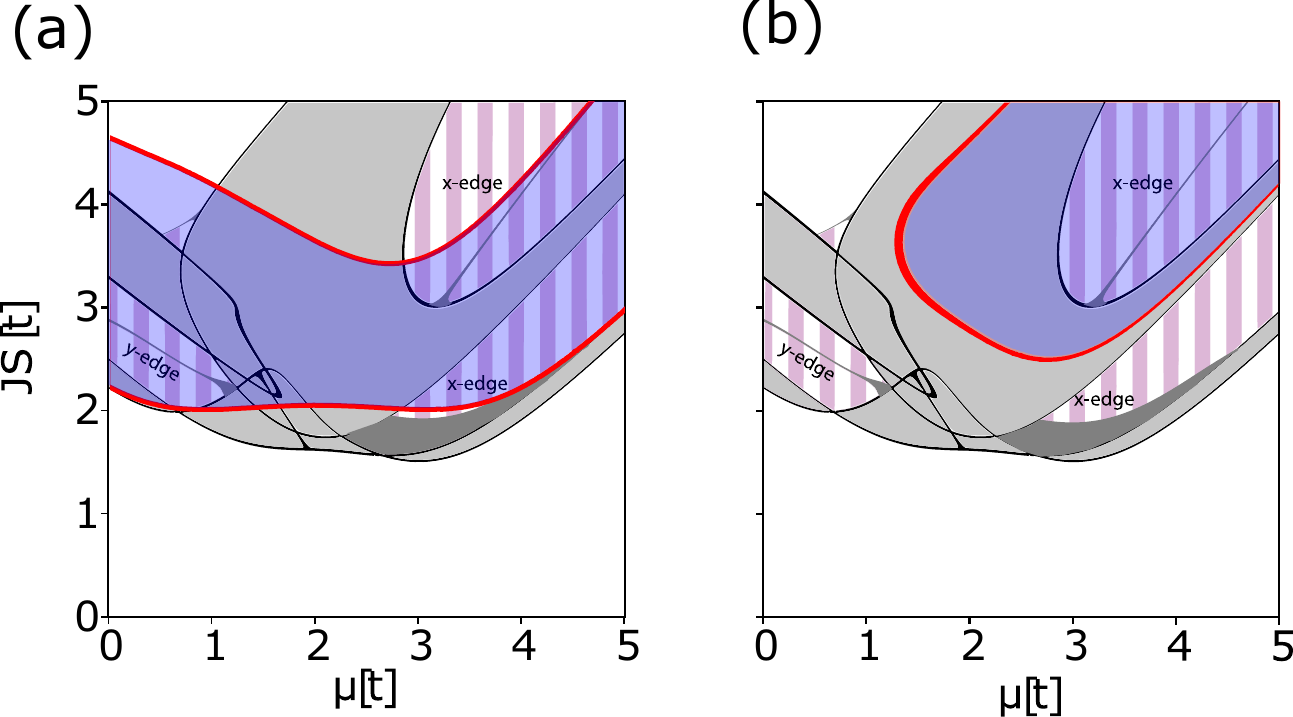}
  \caption{Topological phase diagram of (a) a single adatom chain  [see Fig.~\ref{fig:single}(a)] and (b) a double adatom chain [see Fig.~\ref{fig:single}(b)]
overlaid with the phase diagram of Fig.~\ref{fig:1}(a). Topological phases of the single chain and double chains are denoted by purple regions.}
  \label{fig:overlay}
\end{figure}

The microscopic mechanism underlying the HOTSC phases can be understood by considering the phase diagrams of isolated single and double adatom chains, as shown in Figs.~\ref{fig:single}(a) and (b), respectively. By overlaying the phase diagram of the single chain [see Fig.~\ref{fig:single}(a)] and the double chain [see Fig.~\ref{fig:single}(b)] with the phase diagram in Fig.~\ref{fig:1}(a), as shown in Figs.~\ref{fig:overlay}(a) and (b), respectively,
we find that only the single adatom chain is topological in regions I - III, while both the single and double-chains are topological in region IV. This suggests that the single chain is the fundamental building block underlying the HOTSC phases in regions I-III, while the HOTSC phases in region IV may depend on the interplay between the topological phases of both the single and double chains. in the following, we will therefore consider these two cases separately.

\begin{figure}[h!]
\centering
\includegraphics[width=\linewidth]{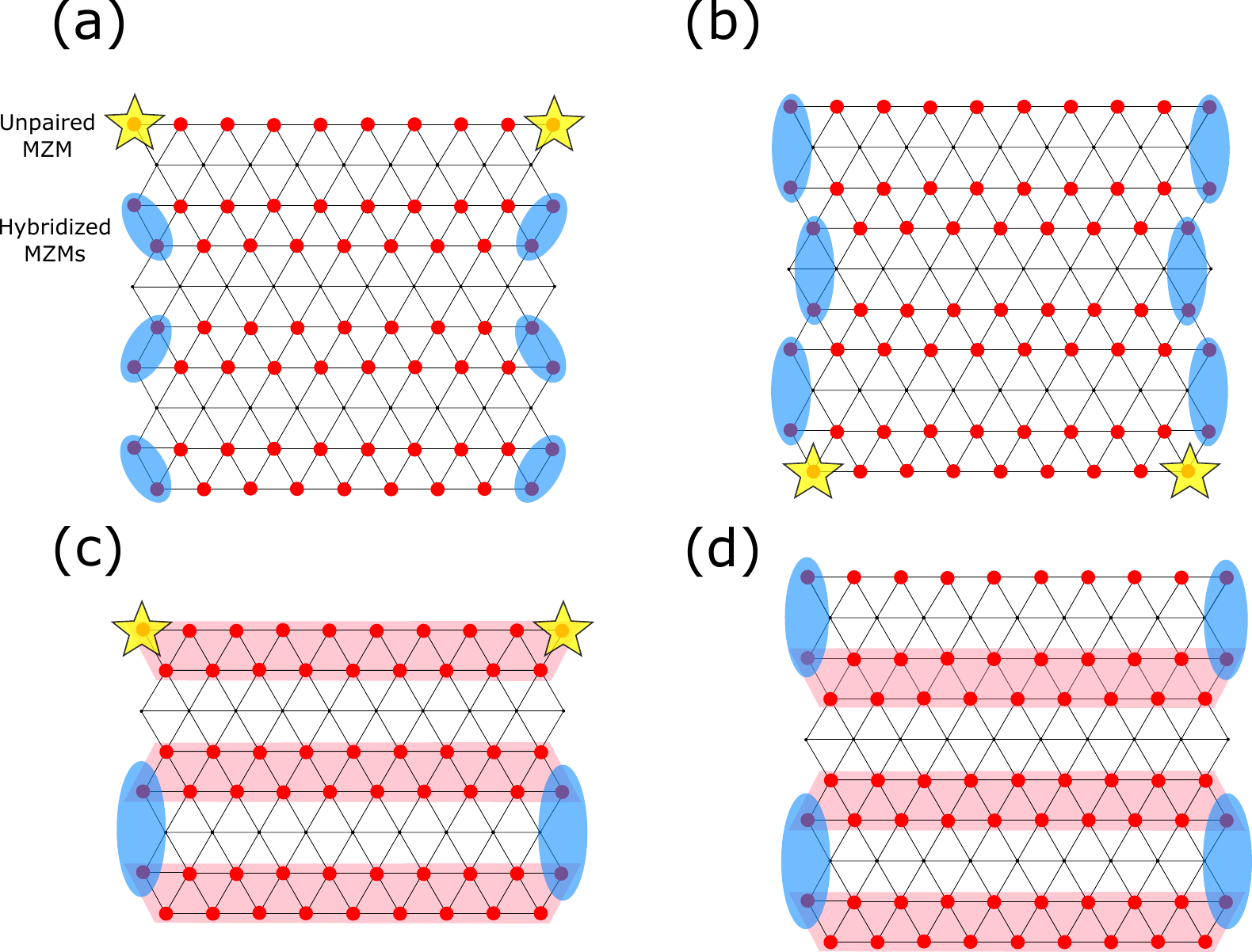}
  \caption{Different ways to pair up the MZMs at the ends of adatom chains when they are in the topological phase. (a) Single chains are topological, and MZMs hybridize across vertical intra-pair couplings, (b) Single chains are topological, and MZMs hybridize across inter-pair couplings, (c) Double chains are topological, and their MZMs hybridize when two double chains are pointing at each other, (d) Single and double chains are topological, so an MZM from a double chain can pair with an MZM from a single chain}
  \label{fig:3}
\end{figure}

\subsubsection{HOTSC phases in regions I-III: the single adatom chain building block}

The origin of the HOTSC phase in regions I-III of the phase diagram can be understood by considering an MSH island with $x$- and $y$-edges built from topological single chains (which themselves extend along the $x$-axis). Each of the single topological chains exhibits a single MZM at their end. Stacking these chains along the $y$-axis can generate the different terminations shown in Figs.~\ref{fig:2}(a)-(d). The corresponding phase diagrams shown in Fig.~\ref{fig:2}(e)-(h), reveal that the HOTSC phases in regions I-III are sensitive only to whether the $x$-edge boundary terminates at a single or a double chain, and is independent of whether the final row points in or out.

To understand these regions of the phase diagrams let us first consider a system with a single chain termination, as depicted in Figs.~\ref{fig:2}(a) and (b). Region I in the corresponding phase diagrams of Figs.~\ref{fig:2}(e) and (f) possesses one trivial and one HOTSC phase that are separated by a phase transition at which the gap along the $y$-edge closes (we refer to this as a $y$-edge transition). The HOTSC phase in this region arises from an intra-pair coupling (i.e., the coupling between the two chains located on adjacent rows) that is stronger than the inter-pair coupling, i.e., the coupling between chains that are separated by an empty row.  The two MZMs within each pair of chains thus strongly hybridize and annihilate (i.e., they move into the bulk continuum). Consequently, only the MZMs belonging to the ends of the single chains at the top and bottom of the MSH island remain (as they do not couple to any other MZMs) and persist as Majorana corner modes, as schematically shown in Fig.~\ref{fig:3}(a). In contrast, in the trivial phase, the relative strength of the intra- and inter-pair couplings are reversed, such that the MZMs of chains separated by an empty row hybridize, and fuse into a bulk state, as schematically shown in Fig.~\ref{fig:3}(b), thus leaving no unpaired MZMs and rendering the system trivial. Finally, at the $y$-edge phase transition separating the trivial and HOTSC phases, the intra- and inter-chain couplings become equal and the MZMs from each of the individual chains couple along the $y$-edge, resulting in the formation of a gapless dispersive mode.

Following this reasoning, it becomes immediately clear that the trivial and topological phases in region I are interchanged between MSH islands with a single chain and a double-chain termination. This can easily be seen from the schematic rendering of the MSH systems in Figs.~\ref{fig:3}(a) and (b): when an additional (topological) chain is added to the top of the MSH system in Fig.~\ref{fig:3}(a), and the termination is thus changed to a double-chain termination, the MZMs of the original single chain and the added chain hybridize, thus rendering the system trivial.
On the other hand, when a single chain is added to the top of the MSH system in Fig.~\ref{fig:3}(b), this chain adds unpaired MZMs that are located in the corners of the MSH system, thus rendering it in the HOTSC phase.  We note that this result is a general feature of a boundary-obstructed extrinsic phase: the topological and trivial phases exchange places in the phase diagram when the edge termination is changed, but the distinction between the two phases survives~\cite{Khalaf2021}.

The mechanism that generates the HOTSC phases in regions II and III is similar to that in region I. However, while region I is divided into two regions where either the intra-chain coupling is stronger than the inter-chain coupling, or vice versa, in regions II and III the inter-pair coupling is always stronger than the intra-pair coupling. Moreover, in contrast to region I, the phase transitions between HOTSC and trivial phases in regions II and III are accompanied by a gap closing along the $x$-edge (we refer to these as $x$-edge transitions). These transitions occur when the single chains themselves undergo a topological phase transition, and thus we expect that the $x$-edge transition lines in the phase diagram should approximately coincide with the phase transition lines of the isolated single chains, as shown in Fig.~\ref{fig:overlay}(b). In region III, this $x$-edge transition occurs over an extended region of the phase diagram (i.e., it is not simply a one-dimensional line). This extended region, however, does not represent a topological phase as the $x$-edge remains gapless simply because of an accidental crossing of trivial bands at the Fermi level.

\subsubsection{Competing HOTSC phase in region IV: interplay of single and double-chains as building blocks}

We next consider the origin of the HOTSC phase in region IV of the phase diagram, a region in which both the isolated single \emph{and} double adatom chains can be topological, as shown in Fig.~\ref{fig:overlay}. To understand the competition and interplay between the two types of non-trivial chains we begin by considering MSH islands with a double chain termination along the $x$-edge. For such a system, each double chain takes the form of a long trapezoid that is stacked in an alternating manner as depicted in Fig.~\ref{fig:3}(c). Moreover, as follows from the LDOS shown in the inset of Fig.~\ref{fig:single}(b), the MZMs of the topological double chains are localized at the sharp outermost corners of this trapezoid. When the long sides of two such trapezoids face each other, the MZMs hybridize strongly, thus annihilating such that the composite system forms a trivial state, as illustrated for the bulk double-chains in Fig.~\ref{fig:3}(c). Thus, when the MSH island is terminated with a trapezoidal double chain, such that the long side of the trapezoid is outward facing [thus realizing a DO termination as shown in Fig.~\ref{fig:2}(c)], isolated MZMs remain at the corners of the system [as shown in Fig.~\ref{fig:3}(c)], thus reflecting the existence of the HOTSC phase, as shown in region IV of the phase diagram in Fig.~\ref{fig:2}(g).

In contrast, if the $x$-edge termination of the MSH system is of the DI type [see Fig.~\ref{fig:2}(d)], all MZMs of the double chain blocks hybridize, yielding a trivial phase in region IV, in agreement with the results shown in Fig.~\ref{fig:2}(h). Thus,  for the DO and DI terminations, the double-chain forms the topological building block and its alternating stacking produces staggered inter-double chain couplings that generate the HOTSC phase. However, unlike the single-chain mechanism in regions I-III, both the stacked structure of the adatom chains as well as the triangular geometry of the lattice (giving rise to trapezoidal double chain units) are required to generate the HOTSC phase.

Finally, we discuss the existence of the HOTSC phase in region IV for MSH systems with SO and SI terminations. This case is the most complicated as it involves the interplay of both the single and double chain units. We begin by noting that MSH islands with
SO and SI terminations can be constructed by adding a single chain to the $x$-edges of a system with DO and DI terminations, respectively. As such, the topology of the resulting systems with SO and SI terminations can be understand by considering the interaction between the double chain termination and the added single chain. For example, an MSH island with a DO termination is in a HOTSC phase in region IV of the phase diagram, thus possessing Majorana corner modes [see Fig.~\ref{fig:3}(c)]. If we add a single chain (transforming the system into one with an SO termination) and that chain is trivial, the original DO termination Majorana corner modes will persist and the system will remain in the HOTSC phase. On the other hand, if the added chain is topological, then the additional MZMs will hybridize with the existing corner MZMs and render the system trivial, as shown in Fig.~\ref{fig:3}(d).

Thus, for an SO terminated MSH system, region IV is divided into a trivial and topological region which correspond approximately to those regions where the single chain is topological or trivial, respectively [see Fig.~\ref{fig:overlay}(a)]. Moreover, this argument immediately implies that these two phases are separated by a gap closing along the $x$-edge, corresponding to the topological phase transition of a isolated single chain. Indeed the $x$-edge transition line in region IV corresponds approximately to the phase transition line of the isolated single chain, as shown in Fig.~\ref{fig:overlay}(a). This line of reasoning thus explains the results shown in Fig.~\ref{fig:2}(e) and (f): the single chain is trivial to the left of the $x$-edge transition line, yielding a HOTSC phase for the SO terminated system in this region, while to the right of the transition, it is topological, thus yielding a trivial phase. The situation is just reversed for an SI-terminated MSH system, since the DI-terminated system (in contrast to the DO-terminated system) is trivial in region IV: thus adding a topological (trivial) single chain to the trivial DI-terminated system will yield an SI-terminated system in the HOTSC (trivial) phase.

\subsection{MSH islands with $x$- and $y^\prime$-edges}
\label{sec:xyprime-edges}

\begin{figure}[h!]
\centering
\includegraphics[width=\linewidth]{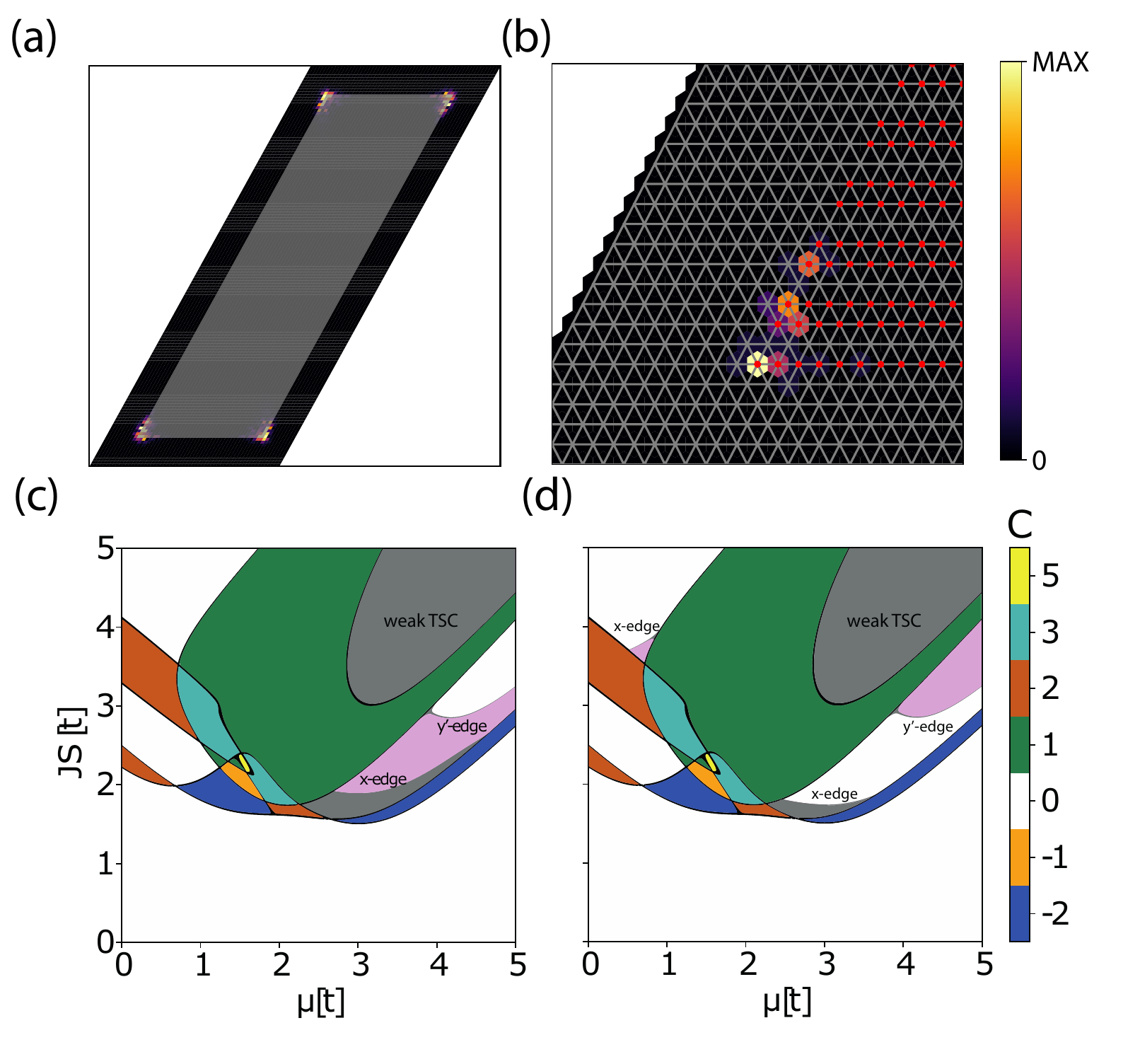}
  \caption{ (a) A parallelogram MSH island shown in light gray, overlaid with the zero-energy LDOS in the HOTSC phase revealing Majorana corner modes at both acute and obtuse angle corners. (b) Zoom-in of (a), showing one acute angle corner. Topological phase diagram for an parallelogram MSH island with (c) a single chain termination, and (d) a double chain termination along the $x$-edge.}
  \label{fig:6}
\end{figure}

As a final example we consider MSH island geometries that possess both $x$- and $y^\prime$-edges. An explicit realization is a parallelogram MSH island, shown in Fig.~\ref{fig:6}(a), whose edges are aligned with the bond-direction of the underlying hexagonal lattice.
In a HOTSC phase, such an MSH islands exhibits Majorana corners modes at both acute and obtues angle corners, as shown in Figs.~\ref{fig:1}(a) and (b). The HOTSC phases of such an island occur in regions II and III of the phase diagram [see Fig.~\ref{fig:1}(b) and Figs.~\ref{fig:6}(c),(d)]. This implies that the mechanism governing the HOTSC phases relies on the topological phase of the isolated single chain, and the relative strength between intra-pair and inter-pair chain couplings that we defined above. Note that, because of the island geometry, the strengths of the intra- and inter-pair couplings in the various regions of the phase diagram are different from MSH islands that have $x$- and $y$-edges. In region II, only the double-chain terminated system exhibits a HOTSC phase, implying that the inter-pair coupling is the dominant one in this region. In region III, MSH systems with single and double-chain terminations exhibit complementary HOTSC phases, and the corresponding topological phase transitions are accompanied by gap closings along the $y^\prime$-edge. Additionally, since this system does not have zigzag edges, the staggered coupling between topological double chains discussed above for the MSH island with $x$- and $y$-edges cannot be realized, and hence no HOTSC phases are found in region IV. Rather, this region exhibits a weak TSC phase ~\cite{seroussi2014topological}, implying a near uniform inter-pair and intra-pair coupling along the $y^\prime$-edge. This mechanism for the emergence of a weak TSC phase is thus similar to that found for rectangular MSH islands located on an underlying square lattice of the superconductor, as discussed in Ref.~\onlinecite{Wong2023}.

\begin{figure}[t!]
\centering
\includegraphics[width=0.9\linewidth]{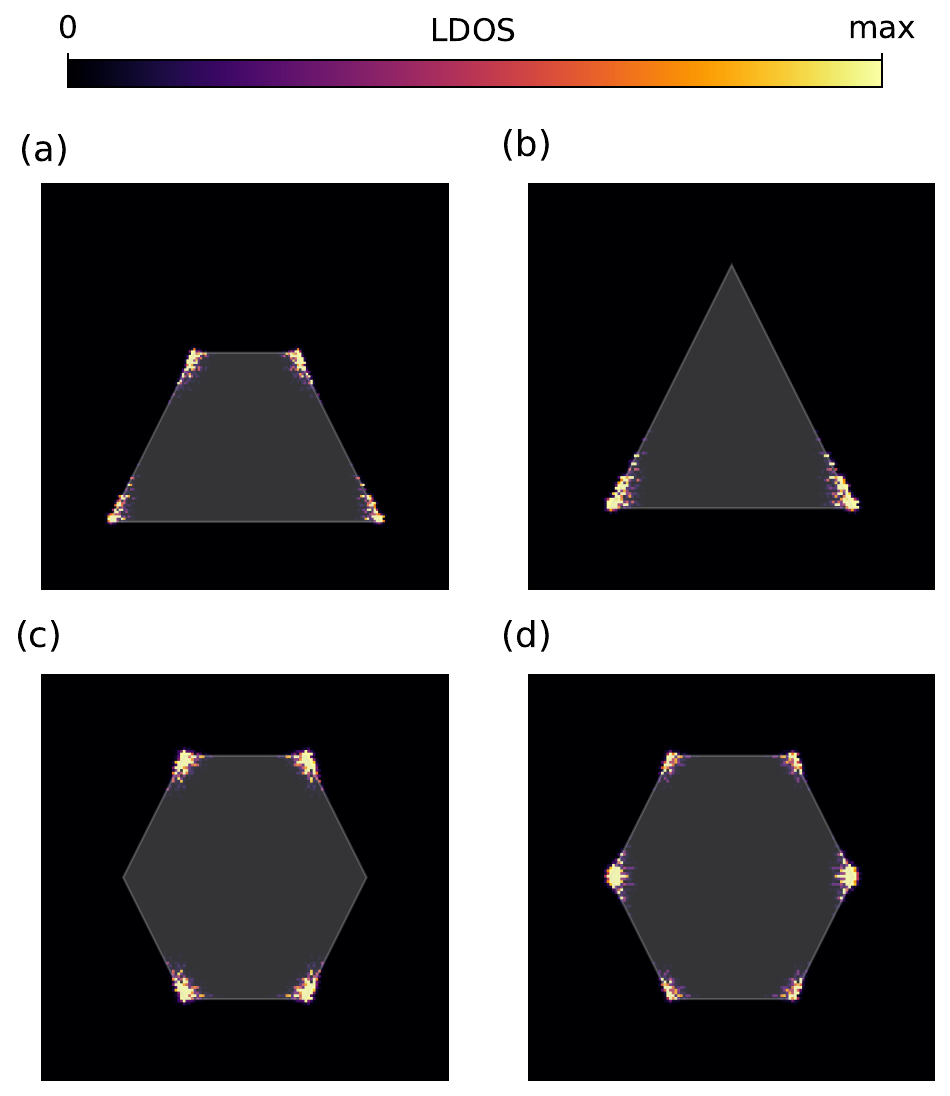}
\caption{ Zero energy LDOS for different MSH island geometries: (a) a trapezoidal island, (b) a triangular island, (c) a hexagonal island formed by stacking of two trapezoids, and (d) a hexagonal island with a single adatom chain inserted in the middle.}
  \label{fig:7}
\end{figure}

Using a combination of $x$- and $y^\prime$-edges, it is also possible to form triangular, trapezoidal, and hexagonal islands as shown in Fig.~\ref{fig:7}. For MSH islands with trapezoidal geometries, the corners of the island are identical to corners of the parallelogram, where $x$- and $y^\prime$-edges meet. Hence, these MSH islands possess the same phase diagrams as those for the parallelogram island shown in Figs.~\ref{fig:6}(c),(d). We note that a trapezoidal island (see Fig.~\ref{fig:7}(a)) can be transformed into a triangular one (see Fig.~\ref{fig:7}(b)) by adding successively shorter chains to the top $x$-edge. As such, the distance between the two Majorana corner modes on the top of the system is reduced with each successive chain, such that they eventually hybridize and annihilate. Thus, a triangular MSH island, as shown in Fig.~\ref{fig:7}(b) exhibits only a single set of Majorana corner modes at the bottom edge, but none at the top corner. At first glance it may appear that this arrangement of MZMs violates the $C_3$ symmetry of the triangular lattice, but this symmetry is already broken by the adatom decoration and therefore does not preclude such arrangements.

The hexagonal island shown in Fig. \ref{fig:7}(c) exhibits two different types of corners: the corners at the top and bottom of the hexagon are formed when $x$- and $y^\prime$-edges meet, and thus again exhibit Majorana Corner modes. In contrast, the middle corners are formed by two $y^\prime$-edges, and thus do not exhibit MZMs. Indeed, such a hexagon can be formed by combining two oppositely oriented trapezoids, which lays plain the reason for the absence of MZMs on the middle corners: the MZMs of the sharp corners of the two trapezoids strongly hybridize at these corners and thus annihilate.
However, one can construct a modified, nearly hexagonal island that possesses MZMs at the middle corners by inserting an additional single adatom chain hosting MZMs in the center of the hexagon, as shown in Fig.~\ref{fig:7}(d).

\section{Discussion}

Our results show that MSH systems with a triangular (superconducting) surface lattice, which are experimentally relevant for the engineering of MSH systems \cite{Palacio-Morales2019}, exhibit a richer HOTSC phase diagram than was previously found for square lattices~\cite{Wong2023}. These HOTSC phases are extrinsic in nature, and thus sensitively depend on the edge termination of an MSH island. For an MSH island with $x$- and $y$-edges, we identified four different edge terminations, labelled SI, SO, DI, and DO. The qualitatively new feature in triangular lattices is that the existence of HOTSC phases depends on the system's termination along the $x$-edge as well as along the $y$-edge. The latter arises from the fact that on a triangular lattice, both the single and double chain are basic building blocks for the MSH system, while on a square lattice, it is only the former. It is the competition and interplay between the topological nature of the single and double chain blocks that gives rise to HOTSC phases that are sensitive not only to the $x$-edge, but also to the $y$-edge termination. The sensitivity of extrinsic HOTSC phases to the edge terminations also results in a phase diagram for an MSH island with $x$- and $y^\prime$-edges that is qualitatively different from that with $x$- and $y$-edges. More generally, we find that the existence of Majorana corner modes can be custom-designed by creating corners in MSH islands made by the intersection of different edges, such as in a MSH triangle, which possesses corners made of $x$- and $y^\prime$-edges as well as of two $y^\prime$-edges. Our results demonstrate that MSH systems with experimentally relevant triangular lattice structure possess a plethora of HOTSC phases, which opens unprecedented possibilities for the quantum engineering of HOTSC phases and the patterning of MZMs by using atomic manipulation techniques to alter edge terminations.\\

\noindent {\bf ACKNOWLEDGEMENTS}\\
We would like to thanks R. Wiesendanger for stimulating discussions. This study was supported by the Center for Quantum Sensing and Quantum Materials, an Energy Frontier Research Center funded by the U. S. Department of Energy, Office of Science, Basic Energy Sciences under Award DE-SC0021238. \\

\noindent {\bf AUTHOR CONTRIBUTIONS} \\
B.B., T.L.H., and D.K.M. conceived and supervised the project. K.H.W., J.G, M.R.H, and A.M. performed the theoretical calculations. All authors discussed the results. J.G., M.R.H., T.L.H., and D.K.M. wrote the
manuscript with contributions from all the authors.\\

\noindent {\bf COMPETING INTERESTS}\\
The authors declare no competing interests.\\

\noindent {\bf DATA AVAILABILITY}\\
Original data are available at https://doi.org/10.5281/zenodo.7749400\\

\noindent {\bf CODE AVAILABILITY}\\
The codes that were employed in this study are available from the authors upon reasonable request.\\


%

\end{document}


\renewcommand{\theequation}{S\arabic{equation}}
\renewcommand{\figurename}{Supplementary Figure}

\title{Competing Higher Order Topological Superconducting Phases in Triangular Lattice Magnet-Superconductor Hybrid Systems}

\author{Ka Ho Wong}\affiliation{\UICPHYS}
\author{Jacopo Gliozzi}\affiliation{\UIUCPHYS}
\author{Mark R. Hirsbrunner}\affiliation{\UIUCPHYS}
\author{Arbaz Malik}\affiliation{\UICPHYS}
\author{Barry Bradlyn}\affiliation{\UIUCPHYS}
\author{Taylor L. Hughes}\affiliation{\UIUCPHYS}
\author{Dirk K. Morr}\affiliation{\UICPHYS}

\maketitle

{\bf Supplementary Note 1: Topological Phase diagram for smaller superconducting order parameter}

\begin{figure}[h]
        \centering
        \includegraphics[width=\linewidth]{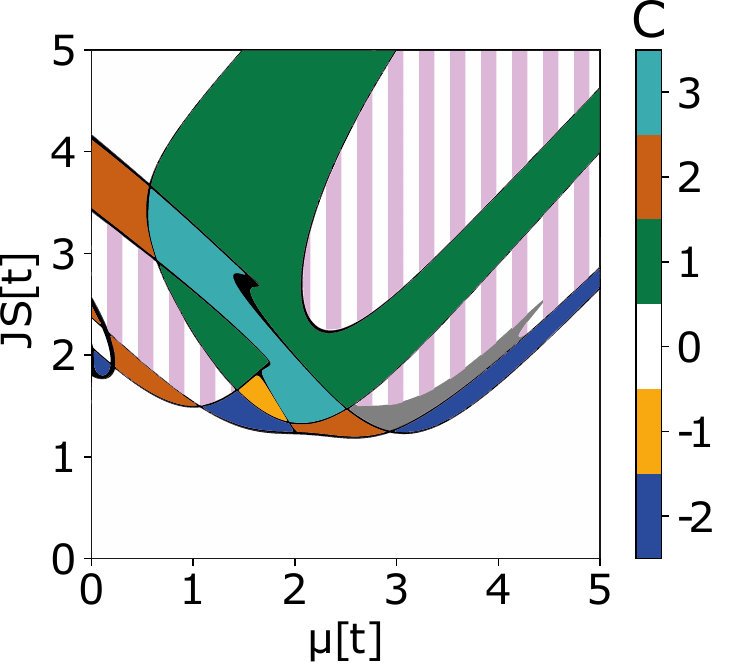}
        \caption{Phase diagram of an MSH island with $x$- and $y$-edges for $\Delta = 0.9t$ and $\alpha=0.6t$. The regions in which HOTSC phases can occur, depending on the choice of edge termination, are indicated by vertical stripes.}
        \label{fig:App1}
\end{figure}

In Supplementary Fig.~\ref{fig:App1} we present the topological phase diagram of an MSH island with $x$- and $y$-edges for values of the superconducting order parameter and Rashba spin-orbit coupling, $\Delta = 0.9t$ and $\alpha=0.6t$, which are different from the ones used in the main text. In particular, $\Delta$ here is smaller than the one used for the results in Fig.~1(b) of the main text. Despite these different parameters, the phase diagrams shown in Supplementary Fig.~\ref{fig:App1} and in Fig.~1(b) of the main text are qualitatively, and to a large extent quantitatively, very similar. In particular, although the boundaries of the strong topological superconducting phases are slightly shifted, the regions in which higher-order superconducting phases can occur persist in the phase diagram. This supports our conclusion that the emergence of HOTSC phases are a general feature of MSH systems with a stacked magnetic structure on an underlying triangular superconducting surface.